\documentstyle[12pt,epsf]{article} \begin{document}

\noindent {\bf Fast X-ray and optical variability of the Black Hole Candidate XTE
J1118+480} \vspace{1\baselineskip}

G.\ Kanbach$^\dagger$, C.\ Straubmeier$^\dagger$, H.C.\ Spruit$^*$,  T.\
Belloni$^\ddagger$

\vspace{1\baselineskip}$^\dagger$Max-Planck-Institut f\"ur Extraterrestrische Physik,
Postfach 1312, 85741 Garching, Germany,

$^*$Max-Planck-Institut f\"ur Astrophysik, Postfach 1317, 85741 Garching,

$^\ddagger$Osservatorio Astronomico di Brera, via Bianchi 46, 23807 Merate (LC), Italy.

\vspace{1\baselineskip}
{\large\it In connection with Nature's publication policy, 
the authors request that the results reported here not be discussed with the
popular press before the publication date}
\vspace{1\baselineskip}

\vspace{1\baselineskip} {\bf 
Black holes become visible when they accrete gas; a close stellar companion
is a common source of the gas. The standard theory ('thin accretion disk')
for this process does not explain some spectacular phenomena, such as
their X-ray variability$^{1}$ and relativistic outflows$^{2}$, indicating some lack of
understanding of the actual physical conditions. Simultaneous observations
at multiple wavelengths can provide strong constraints on the physics. Here
were report simultaneous high-time-resolution X-ray and optical observations
of the transient XTE J1118+480, which show a strong but puzzling correlation
between the emissions. The optical emission rises suddenly following an
increase in the X-ray output, but with a dip 2-5 s in advance of the X-rays.
This result is not easy to understand within the simplest model of the
optical emission, where the light comes from reprocessed X-rays. 
It is probably more consistent with
the earlier suggestion$^{3}$ that the optical light is
cyclosynchrotron emission that originates in a region  about 20,000 km from
the black hole. We propose that the time dependence is evidence for a
relatively slow ($<0.1 c$), magnetically controlled outflow.
}

The X-ray transient XTE J1118+480 (=KV UMa) provided a unique opportunity for simultaneous 
X-ray and optical observations because of its long duration (from January till August 2000), its
proximity, and the lack of obscuration at its position above the galactic plane$^{4-6}$.  The
the orbit of the companion shows$^{5}$ that the central object has a mass $>6$ M$_\odot$. 
We have obtained a total of 2.5 hrs of simultaneous observations distributed over 4 
consecutive nights (July 4 through 7, 2000). The X-rays were strongly variable, in a mode
typical$^{7}$ for black hole candidates (BHC) in the low-hard state, like Cyg X-1. The 
amplitude of the optical variability
was smaller, about 10\% rms on time scales of minutes and shorter. Results of another
simultaneous observation (with RXTE and HST) have been announced earlier$^{8}$.

Correlations between the two time series are hard to identify reliably in the time series
themselves, but are quite unambiguous by cross-correlation of segments as short as 1 min
(Figure 1). An optical response starts within about 30ms following the X-rays, with a
maximum reached after about 0.5 sec. The rapid variability of the optical emission 
(figure 2) excludes intrinsic thermal emission from a cool accretion disk as the source 
of the (variable part of the) optical emission.

The shape of the cross correlation as shown by fig 1 is quite suggestive of a simple
explanation in terms of reprocessing$^{9,10}$:  X-rays from the central regions near the hole
illuminate the upper layers of the surrounding accretion disk, and by heating these cause
optical/UV radiation. The optical response is then easily modeled with an accretion disk inclined
out of plane of the sky. The rapid rise would be due to the relatively short light travel time delay
of photons reprocessed on the earth-facing half of the disk, and the slower decay arising
from the longer path followed by photons reprocessed on the far side. On quantitative
inspection however, this interpretation can be ruled out.  The main problem is that the X-ray
light curve does not have enough variability on time scales of 10-500 ms. With the
observed X-ray autocorrelation (fig 2) a time lag of  order a few seconds can be produced,
but not the observed steep rise in the first few 100 ms. This is also evident from the optical
variability itself. If the optical variation were due to reprocessing of the X-rays, its
autocorrelation should be wider than that of the X-rays. This is the opposite of what is
observed (fig 2). Secondly, there is significant variability in the cross correlation (fig 1, 
bottom), on
time scales as short as 3 minutes. This is hard to explain as a variation in the reprocessing
properties of the disk, since at 0.2-2 lightsec from the center the intrinsic time scales of the 
disk are much longer than 3 minutes. Some variation can be due to changes in the X-ray
autocorrelation function, but quantitatively this again has problem of the lack of sufficient
variability of the X-rays on short time scales. Third, in X-ray binaries where
reprocessing is indicated by independent evidence (such as Bowen fluorescence$^{11}$) the
optical-to-X-ray flux ratio is only a few per cent. This is much less than observed in XTE
J1118+480, 
which had an optical luminosity$^{12}$ around $10^{35}$ erg/s, some 50\% of the X-ray
luminosity. Though our observations address only the (small) variable part of the optical 
emission, the unusually large optical flux itself is not easily explained either as reprocessed
X-rays or as thermal emission from a cool disk.

A curious feature of the cross-correlation is the dip at 2--5s before zero lag (fig 3). The
optical emission appears to decrease before the onset of an X-ray increase. This has been
seen before in the BHC GX 339-4 (the only other source for which  simultaneous X-optical
observations at millisecond time resolution has been reported$^{13}$). This feature poses a
challenge
for any model of the optical emission. Variations in the optical emission preceding the X-rays
suggest a source of variation further out in the disk where (some of the) optical light may be
produced, for example by a change in the mass flow rate. However, the optical signal would
then require a {\em decreasing} mass flux preceding an X-ray increase. Cooling of the X-ray
emitting region by inverse Compton scattering on optical/UV photons$^{3}$ would cause
anticorrelated X-ray and optical variation, but any time delay would be much less than
seconds.

Significant variability in the optical takes place on time scales as short as 100ms (see fig 1,2), hence
the size of the optical emission region is not larger than a light travel distance of $3\,10^9$ cm.
The observed optical brightness $\nu F_\nu=1.5\,10^{-10}$ erg cm$^{-2}$s$^{-1}$ then yields a
minimum brightness temperature of $2\,10^6$K for the inferred distance of the source (1.8 kpc).
The radiation mechanism that most plausibly produces such brightness is cyclosynchrotron
emission (CS) in a strong magnetic field$^{3}$. 
To explain the (variable) delay of the optical emission, we interpret the emission region as the
photosphere (surface of optical depth unity) in a magnetically dominated outflow from the
central regions of the accretion. Modulations in the accretion rate onto the hole, evident in
the X-ray light curve, are assumed to propagate into the outflow as modulations of the flow
velocity and mass flux. Steepened into shocks, these modulations produce the observed
radiation, as in the standard internal shock model of jet emission$^{14,15}$.  The optical
variability reflects the passage of these modulations through the photosphere, and the optical
delay of $\sim 0.5$s is identified with the travel time of the flow to the photosphere from its
origin near the hole. We find that this model is realistic only if the outflow has a rather low
velocity, $v< 30000$ km/s. If the assumed velocity is too large, for example mildly relativistic,
the photosphere is at such a distance (about $10^{10}$ cm) that the magnetic field of $\sim
10^6$G which is required to produce enough optical emission would be unrealistically high.

The outflow proposed here is not the same as that producing the observed radio emission,
which is likely to originate from a less dense but faster flow, possibly
a  mildly relativistic jet$^{16,17}$. Both kinds of outflow may be
present at the same time, concentric to each other. The flow inferred here would then be the
microquasar analog of, for example, the broad-line region outflows$^{18,19}$ of quasars, or
the `anomalous outflows'$^{20}$ of SS433. The CS outflow model is still tentative, however, 
since it does not naturally explain the prominent `precognition dip' at negative lags.

\vfill\eject

\leftline{\bf Refs} 

1 Lewin, W.H.G., van Paradijs, J., van den Heuvel, E.P.J. (eds.) {\it X-ray binaries}.
Cambridge: Cambridge University Press (1995)

2 Mirabel, I.F., Rodr\'{\i}guez, L.F. Sources of Relativistic Jets in the Galaxy. {\it Ann.\
Rev.\ Astron.\ Astrophys.} {\bf 37}, 409--443 (1999)

3 Fabian, A.C., Guilbert, P. W., Motch, C., Ricketts, M.,  Ilovaisky, S.A., Chevalier, C. GX
339-4 -- Cyclotron radiation from an accretion flow.  {\it Astron.\ Astrophys.} {\bf 111},
L9--L12 (1982)

4 Remillard, R.A., Morgan, E., Smith, D., Smith E. XTE 1118+480, {\it IAU Circ.} 7389 (2000)

5 McClintock, J.E., Garcia, M.R., Caldwell, N., Falco, E.E., Garnavich, P.M., Zhao, P. A black
hole of $>6$ M$_\odot$ in the X-ray nova XTE 1118+480. {\it Astrophys.\ J.} {\bf 551},
L147--150 (2001)

6 Wagner, R.M., Foltz, C.B., Shahbaz, T., Casares, J., Charles, P.A., Starrfield, S.G., Hewett,
P. The halo black-hole X-ray transient XTE 1118+480, astro-ph/0104032 (2001)

7 e.g.: Van der Klis, M. Rapid aperiodic variability in X-ray binaries, in ref$^1$, p.252--300
(1995)

8 Haswell, C.A., Skillman, D. Patterson, J. Hynes, R.I. Cui, W. XTE J1118+480, {\it IAU Circ.}
7427 (2000)

9 O'Brien, K. \& Horne, K. Correlated X-ray and optical variability in X-ray Binaries. in {\it
Rossi2000: Astrophysics with the Rossi X-ray Timing Explorer}.  NASA Goddard Space
Flight Center, Greenbelt, MD USA, p.E91-E93 (2000)

10 Hynes, R.I., O'Brien, K., Horne, K., Chen, W., Haswell, C.A. Echoes from an irradiated
disc in GRO J1655-40. {\it Mon. Not. R. Astron. Soc.} {\bf 299}, L37--L41 (1998)

11 Van Paradijs, J. E.\& Mc Clintock, J. Optical and UV observations of X-ray binaries. In
ref. 1, p. 58--121

12 McClintock, J.E., Haswell, C.A., Garcia M.R., et al. Complete and Simultaneous Spectral
Observations of the Black-Hole X-ray Nova XTE J1118+480, {\it Astrophys.\ J.} {\bf 555},
477--482 (2001)

13 Motch, C., Ilovaisky, S.A., Chevalier, C. Discovery of fast optical activity in the X-ray
source GX 339-4. {\it Astron.\ Astrophys.} {\bf 109}, L1--L4 (1982)

14 Fender, R.P., Hjellming, R. M., Tilanus, R.P.J., Pooley, G.G., Deane, J.R., Ogley, R.N.,
Spencer, R.E. Spectral evidence for a powerful compact jet from XTE J1118+480, {\it Mon.\
Not.\ R.\ Astron.\ Soc.} {\bf 322}, L23--L27 (2001)

15 Markoff, S. Falcke, H., Fender, R. A jet model for the broadband spectrum of XTE
J1118+480. Synchrotron emission from radio to X-rays in the Low/Hard spectral state {\it
Astron. Astrophys.} {\bf 372}, L25--28 (2001).

16 Rees, M.J. The M87 jet - Internal shocks in a plasma beam, {\it Mon.\ Not.\ R.\ Astron.\
Soc.} {\bf 184}, 61--65 (1978)

17 Kaiser, C.R., Sunyaev, R.A., Spruit, H.C. Internal shock model for microquasars, {\it
Astron. Astrophys.} {\bf 356}, 975--988 (2000).

18 Leahy, J. P. Phenomenology of Active Galactic Nuclei, in {\it Astrophysical Discs - An EC
Summer School}, Astronomical Society of the Pacific, Conference series Vol  {\bf 160}, Eds.
J.A.\ Sellwood \& J.\ Goodman, p. 246--250. (1999)

19 Sulentic, J.W., Marziani, P., Dultzin-Hacyan, D. Phenomenology of Broad Emission Lines
in Active Galactic Nuclei, {\it Ann.\ Rev.\ Astron.\ Astrophys.} {\bf 38}, 521--571 (2000)

20 Blundell, K.M., Mioduszewski, A., Muxlow, T.W.B., Podsiadlowski, P., Rupen, M. Images of
an equatorial outflow in SS433, {\it Astrophys.\ J.} in press, astro-ph/0109504 (2001)

21 http://heasarc.gsfc.nasa.gov/docs/xte/

22 Straubmeier, C., \& Kanbach, G., Schrey, F. {\it Exp. Astronomy}, in press,
astro-ph/0109181 (2001)

23 http://observatory.physics.uoc.gr/

24 Revnivtsev, M., Sunyaev, R., Borozdi, K. Discovery of a 0.08 HZ QPO in the black hole
candidate XTE1118+480. {\it Astron.\ Astrophys.} {\bf 361}, L37--L39 (2000)

\vspace{2\baselineskip} We thank the XTE time allocation team and Skinakas observatory
for their support and flexibility in making these observations possible, and F.\ Schrey for
technical support.  We thank Drs.\ A.\ Zdziarski, K.\ Horne and R.\ Hynes for valuable
discussions. This work was done with support from the European Commission (TMR
research network program).

\vfill\eject

\leftline{\bf Captions}

Fig. 1 Cross-correlation of the X-ray and optical time series of XTE J1118 +480, showing onset
of an optical response within 30msec. Positive lag corresponds to delay of optical emission.
Top: Average of all 2.5 hrs of data, bottom:  a selection of 3 consecutive 200s segments of
data, illustrating variability of the correlation. Piece-wise linear fits to 30s averages
have been subtracted from both time series before correlating. The X-ray observations were
made with the Rossi X-ray Timing Explorer$^{21}$, the optical observations with the OPTIMA
photometer$^{22}$ attached to the 1.3m telescope at Mt Skinakas, Crete$^{23}$. The
photometer consists of a cluster of fiber-coupled avalanche photo diodes sensitive from 450 to 
950 nm at a mean efficiency of 50\%. Individual photon arrival times are recorded at 2 $\mu$s
absolute accuracy using a GPS-based clock. Like in other observations of this source$^{24}$,
a weak quasiperiodic oscillation of 0.08--0.1 Hz was present in the X-rays.

\vspace{2\baselineskip}

Fig. 2 Autocorrelations of the X-ray and optical time series, showing that the optical
variability is not due to reprocessing of the X-rays. Time scales less than 70 msec are
present in both the X-ray and optical light curves, but the optical autocorrelation is
significantly narrower than the X-ray autocorrelation. If $f_{\rm x}(t), f_{\rm o}(t)$ are the
X-ray and optical light curves and $g(t)$ the optical response of the disk to an infinitesimally
short spike of X-rays, then $f_{\rm o}=f_{\rm x} * g$, where a $*$ denotes convolution. The
X-ray/optical cross correlation is then $C_{\rm xo}=f_{\rm x}*f_{\rm o}=A_{\rm x}*g$ and
$A_{\rm o}=A_{\rm x}*g*g$, where $A_{\rm x,o}=f_{\rm x,o}*f_{\rm x,o}$ are the X-ray 
and optical autocorrelation functions. In a reprocessing
model, the optical autocorrelation must therefore be broader than $A_{\rm x}$.

\vspace{2\baselineskip}

Fig. 3 Correlation on time scales up to 30s, showing the enigmatic `precognition dip' at negative
lags of 2--5s. In the available models for the radiation from black hole candidates the optical light
is produced either simultaneously with the X-rays, or later, by reprocessing of X-rays. Such
models
do not explain how the optical emission can `know something' about  X-ray emission that comes 
later.

\begin{figure} [htb] 
\mbox{}\hfill \hbox{\epsfxsize=1\hsize\epsffile{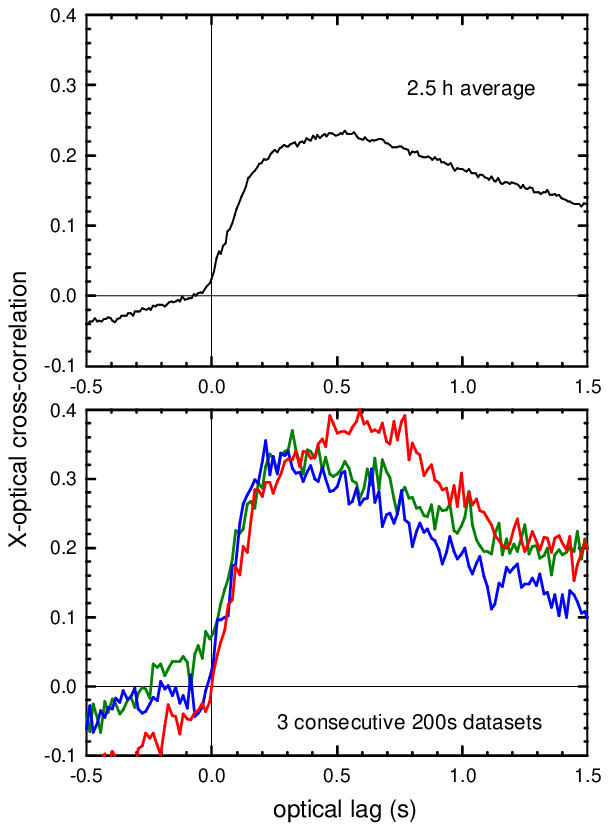}} \hfill\mbox{} 
\caption{ } 
\end{figure}

\begin{figure} [htb] 
\mbox{}\hfill\hbox{\epsfxsize=0.9\hsize\epsffile{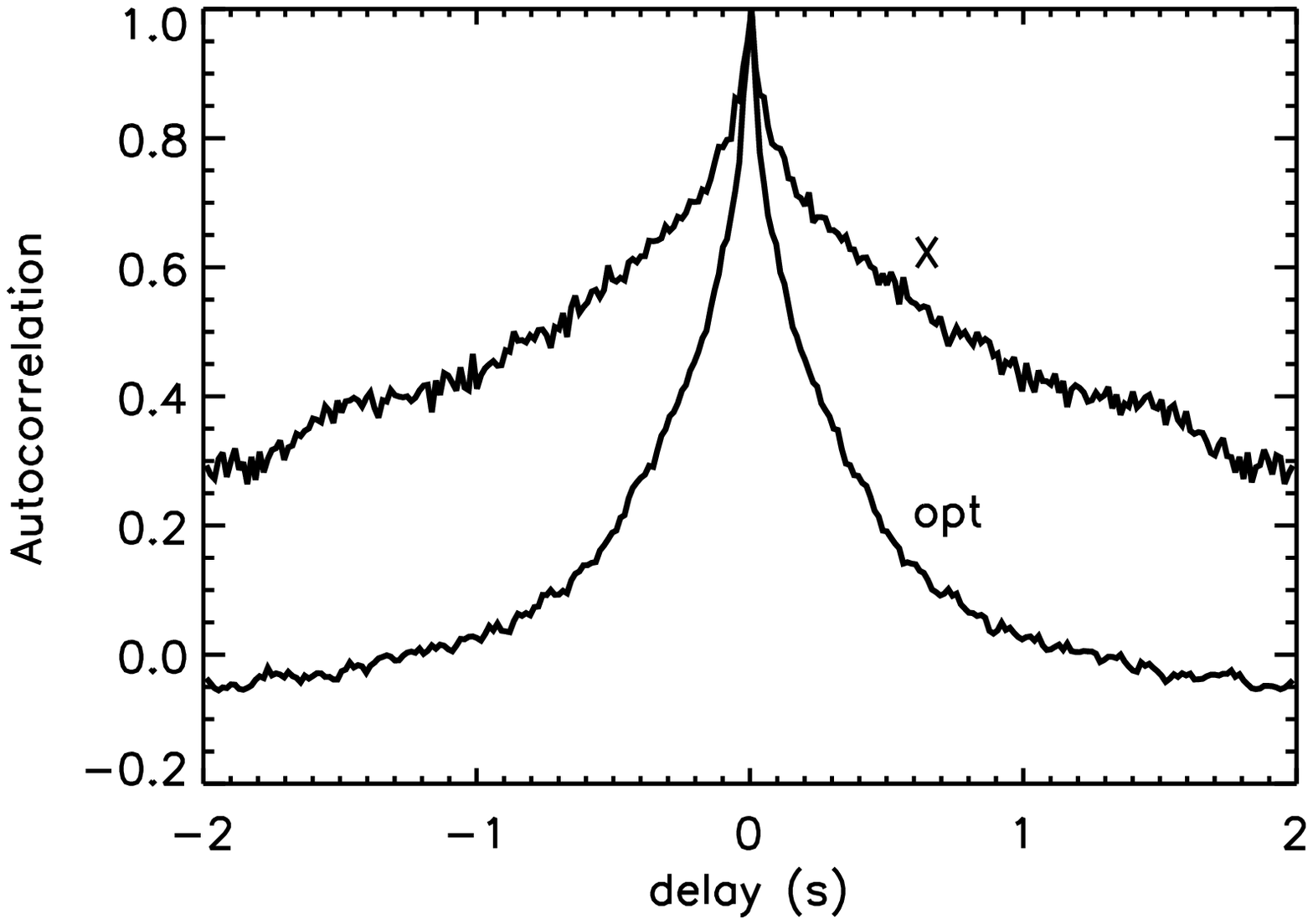}}\hfill\mbox{}
\caption{ } 
\end{figure}

\begin{figure} [htb] 
\mbox{}\hfill\hbox{\epsfxsize=1.0\hsize\epsffile{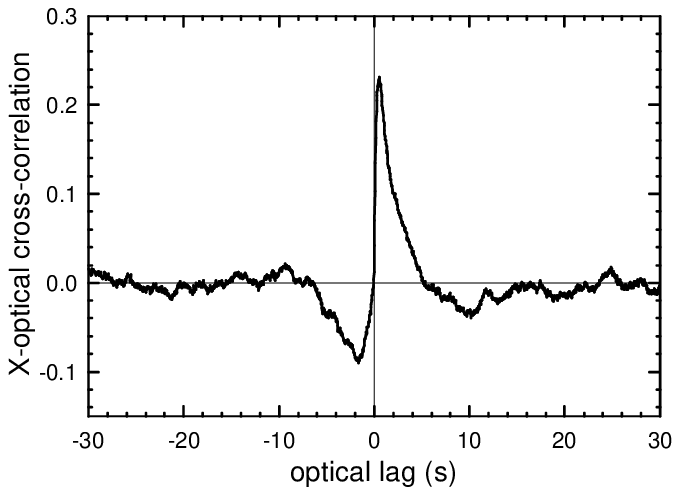}}\hfill\mbox{}
\caption{ } 
\end{figure}

\end{document}